\documentclass[12pt]{article}

\usepackage{amsfonts}
\usepackage{amssymb}
\usepackage{amsthm}
\usepackage[fleqn]{amsmath}
\usepackage[mathcal]{euscript}
\usepackage{mathrsfs}

\newtheorem{corollary}{Corollary}
\newtheorem{proposition}{Proposition}

\renewcommand{\iint}{{\int\!\!\!\int}}

\newcommand{\Ref}[1]{(\ref{#1})}

\newcommand{\vect}[1] {\boldsymbol{{ #1}} }

\newcommand{\pV}{{\vect{p}}}           
\newcommand{\qV}{{\vect{q}}}           

\newcommand{\PV}{{\vect{P}}}           
\newcommand{\QV}{{\vect{Q}}}           

\DeclareMathAlphabet{\mathpzc}{OT1}{pzc}{m}{it}

\newcommand\pzcC{{\mathpzc{C}}}
\newcommand\pzcH{{\mathpzc{H}}}
\newcommand\pzcI{{\mathpzc{I}}}
\newcommand\pzcK{{\mathpzc{K}}}
\newcommand\pzcQ{{\mathpzc{Q}}}

\newcommand{\funcC}{\pzcC}
\newcommand{\funcH}{\pzcH}
\newcommand{\funcI}{\pzcI}
\newcommand{\funcK}{\pzcK}
\newcommand{\funcQ}{\pzcQ}

\newcommand{\oli}[1]{\overline #1 }

\newcommand{\Nset}{\mathbb{N}}
\newcommand{\Rset}{\mathbb{R}}

\newcommand{\Csp}{\mathfrak{C}}
\newcommand{\Hsp}{\mathfrak{H}}
\newcommand{\Lsp}{\mathfrak{L}}

\newcommand{\cE}{{\cal E}}
\newcommand{\cP}{{\cal P}}

\newcommand{\Bose}{\mathscr{Bose}} 
\newcommand{\Coul}{\mathscr{Coul}} 
\newcommand{\Ferm}{\mathscr{Ferm}} 
\newcommand{\Newt}{\mathscr{Newt}} 

\newcommand{\Npow}{{\scriptscriptstyle{N}}}
\newcommand{\npow}{{\scriptscriptstyle{(N-1)}}}
\newcommand{\nnpow}{{\scriptscriptstyle{(N-2)}}}

\newcommand{\tfrhalf}{{\textstyle{\frac{1}{2}}}}
\newcommand{\tfrquarter}{{\textstyle{\frac{1}{4}}}}
\newcommand{\tfrsext}{{\textstyle{\frac{1}{6}}}}

\begin{document}

\title{Monotonicity of quantum ground state energies:\\
	      Bosonic atoms and stars}

\vspace{-0.3cm}
\author{\normalsize \sc{Michael K.-H. Kiessling}\\[-0.1cm]
	\normalsize Department of Mathematics, Rutgers University\\[-0.1cm]
	\normalsize Piscataway NJ 08854, USA}
\vspace{-0.3cm}
\date{$\phantom{nix}$}
\maketitle
\vspace{-1.6cm}

\begin{abstract}
\noindent
	The $N$-dependence of the non-relativistic bosonic ground state energy $\cE^{\Bose}(N)$ 
is studied for quantum $N$-body systems with either Coulomb or Newton interactions. 
	The Coulomb systems are ``bosonic atoms,'' with their nucleus fixed, and it is shown 
that $\cE_{\Coul}^{\Bose}(N)/\cP_{\Coul}(N)$ grows monotonically in $N>1$, where $\cP_{\Coul}(N)=N^2(N-1)$.
	The Newton systems are ``bosonic stars,'' and it is shown that when the Bosons are centrally 
attracted to a fixed gravitational ``grain'' of mass $M>0$, and $N>2$, then 
$\cE_{\Newt}^{\Bose}(N;M)/\cP_{\!\Newt}(N)$ grows monotonically in $N$, where $\cP_{\!\Newt}(N)=N(N-1)(N-2)$; 
in the translation-invariant problem ($M=0$), it is shown that when $N>1$ then 
$\cE_{\Newt}^{\Bose}(N;0)/\cP_{\Coul}(N)$ grows monotonically in $N$, with $\cP_{\Coul}(N)$ from the Coulomb
problem.
	Some applications of the new monotonicity results are discussed.
\end{abstract}
\vfill
\hrule
\smallskip\noindent
{\small 
Typeset in \LaTeX\ by the author.  Revised version of Sept. 10, 2009. 

\smallskip\noindent
\copyright 2009 The author. This preprint may be reproduced for noncommercial purposes.}

\section{Introduction}
\vskip-0.3truecm
\noindent
	While bosonic matter in bulk has been a subject of intense theoretical research over the years, 
spurned on in particular by the recent breakthroughs in creating Bose--Einstein condensates in the laboratory, 
theoretical research into the properties of individual bosonic atoms could seem to always remain of purely 
academic interest, for there are no bosonic electrons in nature. 
	Yet in principle bosonic atoms can exist in nature as we know it, and not just in the
``artificial'' sense described in \cite{KSK}.
	Namely, they can be formed with $N$ bosonic anti-$\alpha$ particles of charge $-2e$ and spin $0$ each 
playing the r\^{o}le of bosonic electrons which are attracted by a nucleus of charge $2eN$, 
conceivably up to $N\approx 46$.
	Both varieties of particles would have to be produced in a laboratory,
the nucleus by ``just'' stripping away all electrons from its associated atom, unless the nucleus 
is itself an $\alpha$ particle which nature supplies through some radioactive materials.
	This latter case yields the simplest (i.e. $N=1$) $\alpha$-bosonic atom, ``alphium,'' 
the $\alpha$ particle analog of protonium,\footnote{Protonium, which is a fermionic atom made of a proton 
		and an anti-proton, recently became an experimental reality \cite{protoniumA,protoniumB}.}
and of the familiar positronium (which would have better been called ``electronium'').
	The $N=2$ $\alpha$-bosonic atom would have a Beryllium nucleus, the stable isotope of which 
(${}^{9}$Be) is a fermion with spin $3/2$; and so on. 
	Since fermionic anti-${}^3$He nuclei have already been produced in heavy ion collisions at CERN 
\cite{AntiHeDREIa,AntiHeDREIb,AntiHeDREIc}, it seems a safe bet to predict that also bosonic anti-$\alpha$ 
particles are going to be produced in the laboratory,\footnote{Should 
		the production of even heavier anti-nuclei become feasible
		some day, then one could also enlist $N>1$ bosonic $\alpha$ particles of charge $2e$ and spin $0$ 
		for playing the r\^{o}le of bosonic electrons orbiting an anti-nucleus of charge $-2eN$.
		Even heavier bosonic atoms are conceivable, e.g. with $N$ Neon nuclei ${}^{20}$Ne of charge $10e$ 
		and spin $0$ each attracted by an anti-nucleus of charge $-10eN$, perhaps up to $N\approx 9$.}
and that research into individual bosonic atoms will take off 
once they can be captured in large enough numbers and made to form bound states with a normal 
nucleus.\footnote{It would be foolish, though, to predict when this will become an experimental reality.}

	Like positronium and protonium, both of which have lifespans of the order of $\mu$-seconds,
also alphium and the heavier $\alpha$-bosonic atoms will be short-lived, but since the involved $\alpha$ and 
anti-$\alpha$ particles are compounds of protons and neutrons, respectively their anti-particles, which 
according to the standard model are themself bound states of up and down quarks, respectively anti-quarks, the 
annihilation modes will be more complicated --- and more interesting --- than those of protonium,\footnote{For 
	an early attempt at calculating effective decay channels of protonium, see \cite{Desai}.} 
and vastly more so than those of positronium.\footnote{The vacuum decay channels of positronium
		are well-known, though not yet completely understood \cite{positronium}.}
	Thus bosonic atoms would also open up a new window for studying the strong interactions at lower energies.

	In the meantime theoretical research into the properties of bosonic atoms remains the only available
venue of inquiry. 
	Following widespread custom \cite{BenguriaLieb}, \cite{Lieb}, \cite{Solovej}, \cite{BachA}, \cite{BachEtAl}, 
\cite{Ruskai},  \cite{BaumgSeiringer}, \cite{Thirring}, in this paper we study the simplified non-relativistic
problem with the atomic nucleus fixed at the origin of a co-ordinate system.
	We will show that $\cE_{\Coul}^{\Bose}(N)/\cP_{\Coul}(N)$ grows monotonically in $N>1$, where 
$\cE_{\Coul}^{\Bose}(N)$ is the bosonic ground state energy of the fixed-nucleus atomic Coulomb Hamiltonian,
and $\cP_{\Coul}(N)=N^2(N-1)$.
	The Galilei-invariant atomic model with a dynamical nucleus is a more tricky $N+1$-body problem 
which we hope to address in the future.

	Our technique of proving monotonicity of the bosonic ground state energy for the atomic Coulomb system easily 
handles also some gravitational Newton system modeling a non-relativistic ``bosonic star.''
	While we have argued that bosonic atoms can in principle exist in nature as we know it and that we expect
them to be produced in laboratories eventually, it is not clear to the author whether bosonic stars will ever be 
more than theoretical speculation.
	In any event, theoretical studies of the bosonic ground state energy for $N$-body Schr\"odinger operators
with gravitational Newton interactions have a long tradition, see
\cite{PostB}, \cite{LevyLeblond}, \cite{HallC,HallD}, \cite{BMR}, \cite{Thirring} (see also 
\cite{LiebYau,Lieb}, \cite{HallLuchaA,HallLuchaB}, \cite{FrankLenzmann} for some 
semi-relativistic models), and so we may as well contribute to it.

	We will first show that when the Bosons are centrally attracted to a fixed gravitational ``grain'' 
of mass $M>0$, and $N>2$, then $\cE_{\Newt}^{\Bose}(N;M)/\cP_{\!\Newt}(N)$ is finite and grows monotonically 
in $N$, where $\cP_{\!\Newt}(N)=N(N-1)(N-2)$.
	Fixing some attracting center --- the nucleus in the atomic and a ``grain'' in the stellar 
case --- is a convenient technical ruse which ensures the existence of a ground state and simplifies the 
mathematics.
	However, while a fixed nucleus is a physically justifiable approximation for normal atoms because
of their large nucleon-to-electron mass ratio, and marginally acceptable for bosonic atoms made 
of anti-$\alpha$ particles bound to a sufficiently large normal nucleus, fixing a gravitational grain 
is entirely artificial and, to the best of the author's knowledge, has not yet attracted much attention in 
the mathematical physics community.
	The problem with a fixed attracting center is of interest chiefly because the limit
$\lim_{M\downarrow 0} \cE_{\Newt}^{\Bose}(N;M)\equiv \cE_{\Newt}^{\Bose}(N;0^+)$
is not only a lower bound for the actual $M=0$ ``ground state'' energy (now read: infimum) $\cE_{\Newt}^{\Bose}(N;0)$ 
of the proper Galilei-invariant $N$-body operator of a bosonic star (obtained by simply setting $M=0$ in the 
operator with fixed attracting center of mass $M$), it can in fact be shown \cite{ReedSimonIV} that 
$\cE_{\Newt}^{\Bose}(N;0^+)$ equals $\cE_{\Newt}^{\Bose}(N;0)$.
	And so, since $\cP_{\!\Newt}(N)$ is independent of $M$, we conclude that also 
$\cE_{\Newt}^{\Bose}(N;0)/\cP_{\!\Newt}(N)$ is finite for $N\geq 3$ and grows monotonically in $N$. 

	Interestingly enough, though, we get a stronger monotonicity result for  $\cE_{\Newt}^{\Bose}(N;0)$ 
by applying our technique directly to the $M=0$ Galilei-invariant $N$-body operator.
	Namely, the familiar reduction of the two-body Hydrogen problem to an effective one-body problem 
with attractive center makes it plain that the translation-invariant $N$-body problem is effectively 
an $N-1$-body problem with attractive center in disguise, obtained by subtracting the energy for the 
degrees of freedom of the $N$-body system's center-of-mass off from the Hamiltonian without changing 
the value of the ``inf'' (though rendering a ``min'' for the so-obtained ``intrinsic Hamiltonian'').
	Our technique applied directly to the reduced Galilei-invariant $M=0$ problem, i.e. the 
intrinsic Hamiltonian, produces the stronger monotonicity law that, when $N>1$ then 
$\cE_{\Newt}^{\Bose}(N;0)/\cP_{\Coul}(N)$ is finite and grows monotonically in $N$, with $\cP_{\Coul}(N)$ as before.
	Note that this monotonicity law implies the monotonicity of 
$\cE_{\Newt}^{\Bose}(N;0)/\cP_{\!\Newt}(N)$ for $N\geq 3$.

	The precise statements of our results are given in section II, their proofs in section III.
	Section IV recalls the Hall-Post inequalities and shows that these are further spin-offs of 
our techniques. 
	We conclude our paper in section V  with an outlook on the fermionic ground state energies.
\section{Results}
%
\subsection{Bosonic atoms with a fixed nucleus}
	Whether one takes $N$ bosonic anti-nuclei of charge $-ze$ each, which repell each other by Coulomb's 
law and are attracted to a nucleus of charge $Nze$ by its electrical Coulomb field, or $N$ bosonic charges 
$ze$ in the field of an anti-nucleus of charge $-Nze$,  with $z\in\Nset$, when the (anti-)nucleus is fixed 
at the origin the non-relativistic $N$-body Hamiltonian for such a bosonic atom in either case is given by 
the formal Schr\"odinger operator 
\begin{equation} 
H^{(N)}_{\Coul}
\equiv
\sum_{1\leq k \leq N} \left({{\frac{1}{2m}}} |\pV_k|^2 -  Nz^2e^2{{\frac{1}{|\qV_k|}}}\right)
+
\sum\sum_{\hskip-.7truecm 1\leq j<k\leq N} z^2e^2{{\frac{1}{|\qV_k-\qV_j|}}},
\label{HamiltonianATOM}
\end{equation}	
where the subscript ${}_{\Coul}$ indicates the electrical Coulomb interactions, and $m$ is the Newtonian 
inertial mass of each of the $N$ particles.
	In \Ref{HamiltonianATOM}, $\pV_k=-i\hbar\nabla_k$ is the familiar momentum operator canonically 
dual to the $k$-th component of the configuration space position operator $\qV_k\in\Rset^3$.
	The formal operator $H^{(N)}_{\Coul}$ is densely defined on 
$\Csp^\infty_0(\Rset^3)\cap \Lsp^2(\Rset^{3N})$.

	As self-adjoint extension we take its Friedrichs extension, also denoted by $H^{(N)}_{\Coul}$,
which is a permutation-symmetric, self-adjoint operator with form domain given by the $N$-fold tensor product 
$D_\funcQ^{(N)}\equiv \Hsp^1(\Rset^3)\otimes \cdots\otimes \Hsp^1(\Rset^3)\subset\Lsp^2(\Rset^{3N})$.
	The quadratic form associated to the operator $H^{(N)}_{\Coul}$ is 
\begin{equation}
\funcQ^{(N)}_{\Coul}(\psi^{(\Npow)}) 
=
{\textstyle{\frac{\hbar^2}{2m}}} \funcK^{(N)}(\psi^{(\Npow)}) -Nz^2e^2\,\funcC^{(N)}(\psi^{(\Npow)}) +z^2e^2\,\funcI^{(N)}(\psi^{(\Npow)}),
\label{QformCOUL}
\end{equation}	

\vskip-0.1truecm
\noindent
where (with integrals extending over $\Rset^{3N}$)
\begin{eqnarray} 
\funcK^{(N)}(\psi^{(\Npow)})
\hskip-0.6truecm&& =
\int  \sum_{1\leq k\leq N}
 |\nabla_k\psi^{(\Npow)}|^2 d^{^{3N}}\!\!\!q,
\label{Kform}\\
\funcC^{(N)}(\psi^{(\Npow)}) 
\hskip-0.6truecm&& =
 \int \sum_{1\leq k\leq N} {{\frac{1} {|\qV_k|}}}|\psi^{(\Npow)}|^2  d^{^{3N}}\!\!\!q,
\label{Cform}\\
\funcI^{(N)}(\psi^{(\Npow)}) 
\hskip-0.6truecm&& =
 \int \sum\sum_{\hskip-.7truecm  1 \leq k < l \leq N} 
\frac{1}{|\qV_k-\qV_l|} |\psi^{(\Npow)}|^2  d^{^{3N}}\!\!\!q.
\label{Iform}
\end{eqnarray}
	The bosonic ground state energy of $H^{(N)}_{\Coul}$ is defined by
\begin{equation} 
  \cE_{\Coul}^{\Bose}(N) 
\equiv
\min\left\{\funcQ^{(N)}_{\Coul}(\psi^{(\Npow)}) \, \Big| \,
	\psi^{(\Npow)}\in D_\funcQ^{(N)} \;;\; \|\psi^{(\Npow)}\|_{\Lsp^2(\Rset^{3N})}=1 \right\}.
\label{qmEnullCOULOMB}
\end{equation}	
	It is well known, e.g. \cite{Thirring}, that a minimizing ground state 
$\psi^{(\Npow)}_{\Bose}$ for \Ref{qmEnullCOULOMB} exists, and that by the permutation symmetry of 
$H^{(N)}_{\Coul}$ the minimizer is permutation symmetric, too, hence ``bosonic.''
	The variational problem \Ref{qmEnullCOULOMB} has been studied in \cite{BenguriaLieb}, \cite{Lieb},
\cite{Solovej}, \cite{BachA}, \cite{BachEtAl}, \cite{Ruskai},  \cite{BaumgSeiringer}, \cite{Thirring};
yet the following monotonicity result for $\cE_{\Coul}^{\Bose}(N)$ seems new.
\begin{proposition}
\label{prop:monoUPa}
	Let $\cE_{\Coul}^{\Bose}(N)$ denote the bosonic ground state energies defined in \Ref{qmEnullCOULOMB}, 
and let $\cP_{\Coul}(N)= N^2(N-1)$.
	Then for $N\geq 2$ the ratio $\cE_{\Coul}^{\Bose}(N)/\cP_{\Coul}(N)$ is finite and grows monotonically in $N$.
\end{proposition}

	Proposition \ref{prop:monoUPa} has some interesting spin-offs.
\begin{corollary}
\label{coro:lowerEofNbound}
	For $N>1$ we have
\begin{equation}
	\cE_{\Coul}^{\Bose}(N) \geq \cE_{\Coul}^{\Bose}(2)\tfrquarter N^3(1-N^{-1}).
\label{lowerEofNbound}
\end{equation}
\end{corollary}
	The  lower bound \Ref{lowerEofNbound} on $\cE_{\Coul}^{\Bose}(N)$ 
is sharp for $N=2$, but certainly not optimal for large $N$.
	The Helium-type ground state energy $\cE_{\Coul}^{\Bose}(2)$ can be calculated, not exactly, but 
approximately with high precision using the method of Hylleraas \cite{Hylleraas}.
	The bound \Ref{lowerEofNbound} may be compared with the bound obtained by setting $k=1$ and $z_1=N$ 
in formula (6.1) in \cite{Lieb}, which reads
\begin{equation}
	\cE_{\Coul}^{\Bose}(N) \geq - (\mathrm{const}.) N^3(1+N^{-4/3}).
\label{lowerEofNboundLIEB}
\end{equation}
	Both bounds have the same leading order power in $N$; if ``(const.)'' in \Ref{lowerEofNboundLIEB}
is $\geq - \tfrquarter \cE_{\Coul}^{\Bose}(2)$, then \Ref{lowerEofNbound} improves over \Ref{lowerEofNboundLIEB} 
for all $N$, but if ``(const.)''$< - \tfrquarter \cE_{\Coul}^{\Bose}(2)$, then \Ref{lowerEofNboundLIEB} beats 
\Ref{lowerEofNbound} for $N>N_*$, with $N_*$ depending on ``(const.).'' 
	Of course, since \Ref{lowerEofNbound} is sharp for $N=2$, ``(const.)'' cannot be smaller than 
$- \cE_{\Coul}^{\Bose}(2)/(8+2^{5/3})$, and if ``(const.)'' equals this value then \Ref{lowerEofNboundLIEB} 
beats \Ref{lowerEofNbound} for all $N>2$. 

	To state our second spin-off of Proposition \ref{prop:monoUPa} we recall that an upper bound to the 
bosonic ground state energy $\cE_{\Coul}^{\Bose}(N)$ is obtained by estimating $\funcQ^{(N)}_{\Coul}(\psi^{(\Npow)})$
from above with the help of a convenient trial wave function
$\psi^{(\Npow)}\equiv\phi^{\otimes \Npow}\in D^{(N)}_Q$, with $\phi\in\Hsp^1(\Rset^3)$.
	We have $\funcQ^{(N)}_{\Coul}(\phi^{\otimes\Npow})=\funcH^{(N)}_{\Coul}(\phi)$, with
\begin{equation}	
\funcH^{(N)}_{\Coul}(\phi) 
=
N{\textstyle{\frac{\hbar^2}{2m}}} \funcK^{(1)}(\phi) -N^2z^2e^2\, \funcC^{(1)}(\phi) +
N(N-1)z^2e^2\tfrhalf\, \funcI^{(2)}(\phi\otimes\phi)
\label{HartreeCoul}
\end{equation}	
a Hartree functional.
	Setting $\phi(\qV) = N^{3/2}\phi_0(N\qV)$ yields the well-known upper bound $\cE_{\Coul}^{\Bose}(N)\leq -C_0N^3$.
	Pairing it with Proposition \ref{prop:monoUPa} we conclude
\begin{corollary}
\label{coro:limitEdurchNhochDREI}
  The limit $\lim_{N\uparrow\infty}N^{-3}\cE_{\Coul}^{\Bose}(N)$ exists and is non-trivial.
\end{corollary}
	Our arguments do not reveal the nature of such a limit.
	In \cite{BenguriaLieb} it is shown that the limit is given by the minimum of the Hartree functional
\begin{equation}	
\funcH_{\Coul}(\phi) 
=
{\textstyle{\frac{\hbar^2}{2m}}} \funcK^{(1)}(\phi) -z^2e^2\, \funcC^{(1)}(\phi) +
z^2e^2\tfrhalf\, \funcI^{(2)}(\phi\otimes\phi) 
\label{HartreeCoulfunc}
\end{equation}
over normalized $\Hsp^1(\Rset^3)$; see also \cite{Lieb},
\cite{Solovej}, \cite{BachA}, \cite{BaumgSeiringer}, \cite{Thirring}.

\subsection{Bosonic stars with a fixed gravitational center}
	The formal Schr\"odinger operator for a bosonic star with fixed gravitational center reads
\begin{equation} 
H^{(N)}_{\Newt\!,M}
\equiv
\sum_{1\leq k \leq N} \left({{\frac{1}{2m}}} |\pV_k|^2 -  GMm{{\frac{1}{|\qV_k|}}}\right)
-
\sum\sum_{\hskip-.7truecm 1\leq j<k\leq N} Gm^2{{\frac{1}{|\qV_j-\qV_k|}}},
\label{HamiltonianSTAR}
\end{equation}	
with $M> 0$.
	The operators of the parameter ($M$) family \Ref{HamiltonianSTAR} are densely defined and symmetric on 
$\Csp^\infty_0(\Rset^3)\cap \Lsp^2(\Rset^{3N})$, and 
as self-adjoint extension of \Ref{HamiltonianSTAR} we again take its Friedrichs extension, also denoted by 
$H^{(N)}_{\Newt\!,M}$, a permutation-symmetric, self-adjoint operator with form domain given by the $N$-fold 
tensor product $D_\funcQ^{(N)}\equiv \Hsp^1(\Rset^3)\otimes \cdots\otimes \Hsp^1(\Rset^3)\subset\Lsp^2(\Rset^{3N})$.
	The quadratic form associated to the operator $H^{(N)}_{\Newt\!,M}$ is
\begin{equation}	
\funcQ^{(N)}_{\Newt\!,M}(\psi^{(\Npow)}) 
=
{\textstyle{\frac{\hbar^2}{2m}}} \funcK^{(N)}(\psi^{(\Npow)}) 
-GMm\, \funcC^{(N)}(\psi^{(\Npow)}) - Gm^2\, \funcI^{(N)}(\psi^{(\Npow)}),
\label{QformNEWT}
\end{equation}	
where $\funcK^{(N)}$, $\funcC^{(N)}$, and $\funcI^{(N)}$ are defined in \Ref{Kform}, \Ref{Cform}, and
\Ref{Iform}, respectively.
	The bosonic ground state energy of $H^{(N)}_{\Newt\!,M}$ for $M>0$ is defined by
\begin{equation} 
\cE_{\Newt}^{\Bose}(N;M) 
\equiv
\min\left\{\funcQ^{(N)}_{\Newt\!,M}(\psi^{(\Npow)}) \, \Big| \,
	\psi^{(\Npow)}\in D_\funcQ^{(N)} \;;\; \|\psi^{(\Npow)}\|_{\Lsp^2(\Rset^{3N})}=1 \right\}.
\label{qmEnullNEWTON}
\end{equation}	
	By the permutation symmetry of $H^{(N)}_{\Newt\!,M}$, the minimizer for \Ref{qmEnullNEWTON} with $M>0$, 
denoted $\psi^{(\Npow)}_{\Bose\!,M}$, is permutation symmetric, too, hence ``bosonic.''

	We will show that the bosonic ground state energies $\cE_{\Newt}^{\Bose}(N;M)$ 
exhibit a monotonic dependence on $N$ similar to Proposition \ref{prop:monoUPa}.
\begin{proposition}
\label{prop:monoUPb}
	For $M>0$ let $\cE_{\Newt}^{\Bose}(N;M)$ denote the bosonic ground state energies defined in 
\Ref{qmEnullNEWTON}, and let $\cP_{\!\Newt}(N)= N(N-1)(N-2)$. 
	Then for $N\geq 3$ the ratio $\cE_{\Newt}^{\Bose}(N;M)/\cP_{\!\Newt}(N)$ is finite and grows 
monotonically in $N$.
\end{proposition}

	Also Proposition \ref{prop:monoUPb} has two technical spin-offs. 
	The first one is immediate:
\begin{corollary}
\label{coro:lowerEofNboundM}
	For $N>2$ we have 
\begin{equation}
	\cE_{\Newt}^{\Bose}(N;M) \geq \cE_{\Newt}^{\Bose}(3;M)\tfrsext N^3(1-N^{-1})(1-2N^{-1}).
\label{lowerEofNboundM}
\end{equation}
\end{corollary}
	The lower bound \Ref{lowerEofNboundM} on $\cE_{\Newt}^{\Bose}(N;M)$ is sharp for $N=3$, but 
far from sharp when $N\gg 1$.
	The coefficient $\cE_{\Newt}^{\Bose}(3;M)$ can be estimated from below,\footnote{Incidentally,
		since all $\cE_{\Newt}^{\Bose}(N;0^+)<0$, 
		neither $\cE_{\Newt}^{\Bose}(2;M)$ nor any $\cE_{\Newt}^{\Bose}(N;M)$ for $N>2$
		can be estimated from below uniformly in $M$ by some fixed multiple of the explicitly 
		computable one-body ground state energy $\cE_{\Newt}^{\Bose}(1;M)=-\tfrhalf G^2M^2m^3/\hbar^2$.}
uniformly in $M$, in terms of the two-body ground state energy with central mass $M/2$ and gravitational constant 
$2G$, as follows: $\cE_{\Newt}^{\Bose}(3;M)\geq 3\cE_{2\Newt}^{\Bose}(2;M/2)$, where the notation ``$2\Newt$'' stands
for the replacement of $G$ by $2G$. 
	Neither $\cE_{\Newt}^{\Bose}(3;M)$ nor $\cE_{2\Newt}^{\Bose}(2;M/2)$ are known to be exactly computable, 
but the Helium-type ground state energy $\cE_{2\Newt}^{\Bose}(2;M/2)$ can certainly be computed in very accurate 
approximation by Hylleraas' variational method \cite{Hylleraas}.

	To state our second spin-off of Proposition \ref{prop:monoUPb} we recall that
an upper bound to the bosonic ground state energy $\cE^{\Bose}_{\Newt\!,M}(N)$ 
is obtained by estimating $\funcQ^{(N)}_{\Newt\!,M}(\psi^{(\Npow)})$ from above
with a convenient trial wave function $\psi^{(\Npow)} \equiv\phi^{\otimes \Npow}\in D^{(N)}_Q$, 
with $\phi\in\Hsp^1(\Rset^3)$.
	This gives $\funcQ^{(N)}_{\Newt\!,M}(\phi^{\otimes\Npow})=\funcH^{(N)}_{\Newt\!,M}(\phi)$, where
\begin{equation}	
\funcH^{(N)}_{\Newt\!,M}(\phi) 
  =
N{\textstyle{\frac{\hbar^2}{2m}}}
 \funcK^{(1)}(\phi)-NGMm\,\funcC^{(1)}(\phi)-N(N-1)Gm^2\tfrhalf\,\funcI^{(2)}(\phi\otimes\phi)
\label{HartreeNEWTwM}
\end{equation}	
is a Hartree functional.
	Setting $\phi(\qV) = N^{3/2}\phi_0(N\qV)$, one easily obtains upper bounds on the Hamiltonian ground
state energies which are $\propto -N^3(1+O(1/N))$.
	Pairing such an upper bound with Proposition \ref{prop:monoUPb} we obtain
\begin{corollary}
\label{coro:limitEdurchNhochDREIb}
  The limit $\lim_{N\uparrow\infty}N^{-3}\cE^{\Bose}_{\Newt}(N;M)$ exists and is non-trivial.
\end{corollary}
	Our arguments do not reveal the nature of the limit, yet it is natural to conjecture that
it is given by the minimum of the limiting Hartree functional
\begin{equation}	
\funcH_{\Newt}(\phi) 
  =
{\textstyle{\frac{\hbar^2}{2m}}} \funcK^{(1)}(\phi)- Gm^2\tfrhalf\,\funcI^{(2)}(\phi\otimes\phi)
\label{HartreeNEWT}
\end{equation}	
which does not feature $M$.
	This should be provable along the lines of 
\cite{BenguriaLieb} and \cite{LiebYau}; see also \cite{Lieb}, \cite{Thirring}.
\medskip

\hskip-.1truecm
	We stress that the above stated results hold for any grain's mass $M>0$.

\subsection{Bosonic stars: the Galilei-invariant model}
	When $M=0$ then the ``min'' in \Ref{qmEnullNEWTON} has to be replaced by ``inf.'' 
	Yet, for $M=0$ the Hamiltonian \Ref{HamiltonianSTAR} can be decomposed as
\begin{equation} 
H^{(N)}_{\Newt\!,0}
\equiv
H^{(1)}_{\rm free} + H^{(N-1)}_{\Newt\!,\rm int},
\label{HamiltonianSTARnullSPLIT}
\end{equation}	
where $H^{(1)}_{\rm free}$ is the Hamiltonian of a single non-interacting (free) particle of mass $Nm$,
describing the center-of-mass motion of the $N$-body system,  
\begin{equation} 
H^{(1)}_{\rm free} \equiv {\textstyle{\frac{1}{2Nm}}} |\PV|^2,
\label{HamiltonianFREEm}
\end{equation}	
where $\PV = \sum_{k=1}^N\pV_k$,  while $H^{(N-1)}_{\Newt\!,\rm int}$ is the reduced Hamiltonian for
the remaining degrees of freedom of the $N$-body system, describing the system-intrinsic motions, in
effect an $N-1$-body problem.
	The intrinsic Hamiltonian is written most symmetrically in the vector variables of \Ref{HamiltonianSTAR}, 
viz.
\begin{equation} 
H^{(N-1)}_{\Newt\!,\rm int}
\equiv
\sum\sum_{\hskip-.7truecm 1\leq j<k\leq N}
	\left({{\frac{1}{2Nm}}} |\pV_j-\pV_k|^2  -  Gm^2{{\frac{1}{|\qV_j-\qV_k|}}}\right),
\label{HamiltonianSTARintrinsic}
\end{equation}
see the first (unnumbered) equation on p.382 in \cite{HallPost}, see also Eq.(2.3) in \cite{HallC};
however, only $N-1$ vectors of the set $\{\qV_k\}_{k=1}^N$ are linearly independent ---
in other words, the vectors in $\{\qV_k\}_{k=1}^N$ are linear combinations of $N-1$ basis vectors
in the linear subspace $\{\QV\equiv {\mathbf{0}}\}^\perp\subset\Rset^{3N}$, 
where $\QV = N^{-1}\sum_{k=1}^N\qV_k$ is the position vector of the system's center-of-mass 
canonically conjugate to $\PV$ (up to scaling).
	Orthogonal transformations from $\{\qV_k\}_{k=1}^N$ to $\{\QV\}\cup\{\oli\qV_k\}_{k=1}^{N-1}$ having
unit Jacobian determinant are described in \cite{PostA} and \cite{HallC}, for instance.
	The Friedrichs extension of the intrinsic Hamiltonian \Ref{HamiltonianSTARintrinsic} takes its minimum
on its form domain $\Hsp^1(\Rset^3)\otimes \cdots\otimes \Hsp^1(\Rset^3)\subset\Lsp^2(\Rset^{3(N-1)})$, and
its minimum agrees with the infimum of the full Hamiltonian \Ref{HamiltonianSTAR} for $M=0$.
\begin{proposition}
\label{prop:monoUPc}
	Let $\cE_{\Newt}^{\Bose}(N;0)$ denote the bosonic ground state energies defined in \Ref{qmEnullNEWTON}
with $M=0$ and ``min'' replaced by ``inf.''
	Then for $N\geq 2$ the ratio $\cE_{\Newt}^{\Bose}(N;0)/\cP_{\Coul}(N)$ is finite and grows 
monotonically in $N$.
	Here, $\cP_{\Coul}(N)=N^2(N-1)$ is the same polynomial which occurs in 
the $\!$``atomic''$\!$ Proposition~\ref{prop:monoUPa}.
\end{proposition}
\begin{corollary}
\label{coro:lowerEofNboundNULL}
	For $N>1$ we have 
\begin{equation}
	\cE_{\Newt}^{\Bose}(N;0) 
\geq \cE_{\Newt}^{\Bose}(2;0)\tfrquarter N^3(1-N^{-1}).
\label{lowerEofNboundNULL}
\end{equation}
\end{corollary}
	The inequality \Ref{lowerEofNboundNULL} is sharp for $N=2$, but far from optimal when $N\gg 1$.
	The lower bound \Ref{lowerEofNboundNULL} is in fact known since \cite{PostB}, 
where it was proved with different arguments; see also \cite{HallA}, \cite{HallB}, \cite{BMR}.
	A slightly weaker bound was obtained in \cite{LevyLeblond}, where it was proved that for $N\geq 2$,
\begin{equation}
	\cE_{\Newt}^{\Bose}(N;0) \geq \cE_{\Newt}^{\Bose}(2;0)\tfrhalf N^3(1-N^{-1})^2.
\label{lowerEofNboundNuLL}
\end{equation}
	Also inequality \Ref{lowerEofNboundNuLL} is sharp for $N=2$, but far from optimal when $N\gg 1$.
	We remark that one can explicitly compute $\cE_{\Newt}^{\Bose}(2;0) = -\tfrquarter G^2m^5/\hbar^2$.

	Next we recall that in \cite{HallA} also the mirror-symmetric upper bound 
\begin{equation}
	\cE_{\Newt}^{\Bose}(N;0) \leq -B N^3(1-N^{-1})
\label{upperEofNboundNuLL}
\end{equation}
was proved without invoking a Hartree functional (more explicitly, set $p=-1$ in formula (1.3) in \cite{HallB}).
	Yet, while formula (1.3) in \cite{HallB} does not imply that 
$\cE^{\Bose}_{\Newt}(N;0)/\cP_{\Coul}(N)$ converges as $N\to\infty$, our
Proposition \ref{prop:monoUPc} paired with Hall's upper bound \Ref{upperEofNboundNuLL} gives right away
\begin{corollary}
\label{coro:limitEdurchNhochDREIc}
  The limit $\lim_{N\uparrow\infty}\cE^{\Bose}_{\Newt}(N;0)/\cP_{\Coul}(N)$ exists and is non-trivial.
\end{corollary}
	It is known \cite{Thirring} that the limit is given by the minimum of the Hartree functional
\Ref{HartreeNEWT}.
	Of course, the upper bound $\cE_{\Newt}^{\Bose}(N;0)\leq -C N^3$ with $C$ coming from 
Hartree theory, obtained earlier in \cite{PostB}, \cite{LevyLeblond}, can also be paired with 
Proposition \ref{prop:monoUPc} to yield Corollary \ref{coro:limitEdurchNhochDREIc}.
	
	By the way,  the convergence of the various monotonic increasing sequences in
Propositions \ref{prop:monoUPa}, \ref{prop:monoUPb}, and \ref{prop:monoUPc} follows already from the
negativity of all the ground state energies --- what does not follow, then, is the nontriviality of the limits.
\section{Proofs}
	Our Propositions \ref{prop:monoUPa}, \ref{prop:monoUPb}, and \ref{prop:monoUPc} are   
inspired by a monotonicity result for classical ground state energies proved in \cite{KieRMP} 
and elaborated on in \cite{KieJSPa}.
	The classical proposition also covers Coulomb charges which, instead of being attracted by a 
nucleus, are confined to a sphere or some other compact domain, and then 
$\cE^{cl}_{\Coul}(N)/\cP^{cl}_{\Coul}(N)$ grows monotonically,
where $\cE^{cl}_{\Coul}(N)$ is the classical Coulomb ground state energy and $\cP^{cl}_{\Coul}(N)=N(N-1)$.
	We shall rewrite the functionals of the quantum ground state energy variational principles into
a quasi-classical format and then recycle the classical estimates. 
	Yet the proofs of Propositions \ref{prop:monoUPa}, \ref{prop:monoUPb}, and \ref{prop:monoUPc} do not
just consist of such variants of the classical estimate in \cite{KieRMP,KieJSPa} but also use the virial 
theorem in an essential way; the virial theorem plays no r\^{o}le in the classical proof.
	Incidentally, to apply the virial theorem we need $\cE^{\Bose}(N)$ to be a minimum, not just an
infimum.

\subsection{Proof of Proposition \ref{prop:monoUPa}}
\vskip-.3truecm
\noindent
	We begin by rewriting the quadratic form $\funcQ^{(N)}_{\Coul}(\psi^{(\Npow)})$ into the convenient
format of a quasi-classical expectation functional.
	Recall the physicists' {\it non-unitary} Fourier transform\footnote{This
		differs only by scaling from the conventional unitary Fourier transform.}
of $\psi^{(\Npow)}$,
\begin{equation}
\widehat{\psi}^{(\Npow)}_\hbar(\pV_1,...,\pV_{N}) 
:= \int \psi^{(\Npow)}(\qV_1,...,\qV_{N})e^{- i {\pV}\cdot{\qV}/\hbar} d^{^{3N}}\!\!\!q,
\end{equation}
so that
\begin{equation}
\psi^{(\Npow)}(\qV_1,...,\qV_{N}) = 
\frac{1}{h^{3N}} \int \widehat{\psi}^{(\Npow)}_\hbar(\pV_1,...,\pV_{N}) e^{i {\pV}\cdot{\qV}/\hbar} d^{^{3N}}\!\!\!q,
\end{equation}
where $h=2\pi \hbar$ is Planck's quantum of action. 
	This Fourier transform is a {\it non-isometric} isomorphism of $\Lsp^2(\Rset^{3N})$, so
when $\|\psi^{(\Npow)}\|_{\Lsp^2(\Rset^{3N})}=1$, then
\begin{equation}
\int |\widehat{\psi}^{(\Npow)}_\hbar(\pV_1,...,\pV_{N})|^2 d^{^{3N}}\!\!\!p  = h^{3N}.
\end{equation}
	Clearly, $h^{-3N} |\widehat{\psi}^{(\Npow)}_\hbar|^2 |\psi^{(\Npow)}|^2 \geq 0$, and 
$\iint h^{-3N} |\widehat{\psi}^{(\Npow)}_\hbar|^2 |\psi^{(\Npow)}|^2  d^{^{3N}}\!\!\!\!p d^{^{3N}}\!\!\!\!q = 1$
when $\|\psi^{(\Npow)}\|_{\Lsp^2(\Rset^{3N})}=1$, so we can think of 
$h^{-3N} |\widehat{\psi}^{(\Npow)}_\hbar|^2 |\psi^{(\Npow)}|^2$ as a formal probability density function on the 
$N$-body phase space of points $(\pV_1,...,\pV_{N};\qV_1,...,\qV_{N})\in \Rset^{6N}$.
	With the help of this Fourier transform, and integration by parts, we can rewrite the quadratic 
form \Ref{QformCOUL} into a quasi-classical ensemble average thusly,
\begin{equation}
\funcQ^{(N)}_{\Coul}(\psi^{(\Npow)}) 
=
\iint H^{(N)}_{\Coul}h^{-3N} |\widehat{\psi}^{(\Npow)}_\hbar|^2 |\psi^{(\Npow)}|^2d^{^{3N}}\!\!\!p d^{^{3N}}\!\!\!q
=:\big\langle H^{(N)}_{\Coul}\big\rangle_{\psi^{(\Npow)}}
\label{QformCOULrewrite}
\end{equation}	
where the double integral extends over $\Rset^{6N}$, and $H^{(N)}_{\Coul}(\pV_1,...,\qV_{N})$ now 
is the {\it classical} Hamiltonian with Coulomb interactions, formally also given 
by \Ref{HamiltonianATOM} but now with $\pV_k\in\Rset^3$; we use the same symbol for the Hamiltonian operator 
and its classical counterpart, as the context makes it unambiguously clear which object is meant.

	Using next a familiar trick of Fisher and Ruelle \cite{FisherRuelle} and Dyson and Lenard 
\cite{DysonLenard}, we rewrite the Coulomb Hamiltonian \Ref{HamiltonianATOM} as a double sum,
\begin{equation}
H^{(N)}_{\Coul}
\equiv
\sum\sum_{\hskip-.7truecm  1 \leq k < l \leq N} U^{(N)}_{k,l}, 
\label{HamCOULrewrite}
\end{equation}
where
\begin{equation}
U^{(N)}_{k,l}
:=
{\textstyle{\frac{1}{2m(N-1)}}}\left( |\pV_k|^2 + |\pV_l|^2 \right) 
-  \textstyle{\frac{Nz^2e^2}{N-1}} \left({{\frac{1}{|\qV_k|}}}+{{\frac{1}{|\qV_l|}}}\right)
+z^2e^2{\textstyle{\frac{1}{|\qV_k-\qV_j|}}}.
\label{HamCOULpairTERM}
\end{equation}
	The superscript $^{(N)}$ at $U^{(N)}_{k,l}$ reminds us of the explicit $N$ dependence
exhibited at r.h.s.\Ref{HamCOULpairTERM}.
	With the help of \Ref{HamCOULrewrite} the quadratic form alias expectation functional 
\Ref{QformCOULrewrite} becomes 
\begin{equation}	
\funcQ^{(N)}_{\Coul}(\psi^{(\Npow)})
\equiv
\sum\sum_{\hskip-.7truecm  1 \leq k < l \leq N} \big\langle U^{(N)}_{k,l}\big\rangle_{\psi^{(\Npow)}}.
\label{QformCOULrewriteDOUBLEsum}
\end{equation}	
	The double sum at r.h.s.\Ref{QformCOULrewriteDOUBLEsum} can be represented graph-theoretically as a complete
$N$-graph with vertices numbered $1,...,N$, with a value $\big\langle U^{(N)}_{k,l}\big\rangle_{\psi^{(\Npow)}}$
assigned to the bond between the $k$-th and $l$-th vertex. 
	An elementary graph-theoretical identity used in the classical proof in \cite{KieRMP,KieJSPa} says 
that such a sum over all bonds  in a complete $N$-graph with $N>2$ equals $(N-2)^{-1}$ 
times the sum over all bonds of all its complete $N-1$-subgraphs. 
	So for $N>2$,
\begin{equation}
\sum\sum_{\hskip-.6truecm  1 \leq k < l \leq N}  \big\langle U^{(N)}_{k,l}\big\rangle_{\psi^{(\Npow)}}
=
{\textstyle{\frac{1}{N-2}}}\!\!\sum_{1\leq n \leq N}
\sum\sum_{\hskip-.6truecm \genfrac{}{}{0pt}{}{1\leq k < l\leq N}{ k\neq n\neq l} } 
\big\langle U^{(N)}_{k,l}\big\rangle_{\psi^{(\Npow)}} .
\label{QformCOULgraphID}
\end{equation}	
	Note that \Ref{QformCOULgraphID} holds without any particular symmetry assumption on $\psi^{(\Npow)}$.

	We now start our estimates.
	Writing $\min_{\psi^{(\Npow)}}$ for the minimum over 
the subset of ${D}_\funcQ^{(N)}$ satisfying $\|\psi^{(\Npow)}\|_{\Lsp^2(\Rset^{3N})}=1$, for $N>2$ we find
\begin{eqnarray} 
\cE_{\Coul}^{\Bose}(N)
\hskip-0.6truecm&& =
\min_{\psi^{(\Npow)}}
\sum\sum_{\hskip-.6truecm 1\leq k< l\leq N} \big\langle U^{(N)}_{k,l}\big\rangle_{\psi^{(\Npow)}}
\nonumber\\
&& \geq
{\textstyle{\frac{1}{N-2}}}\!\!\sum_{1\leq n \leq N}
\min_{\psi^{(\Npow)}}
\sum\sum_{\hskip-.6truecm \genfrac{}{}{0pt}{}{1\leq k < l\leq N}{k\neq n\neq l}}
\big\langle U^{(N)}_{k,l}\big\rangle_{\psi^{(\Npow)}} 
\nonumber\\
&& \geq
{\textstyle{\frac{1}{N-2}}}\!\!\sum_{1\leq n \leq N}
\min_{\psi^\npow}
\sum\sum_{\hskip-.6truecm \genfrac{}{}{0pt}{}{1\leq k < l\leq N}{k\neq n\neq l}}
\big\langle U^{(N)}_{k,l}\big\rangle_{\psi^{\npow}} 
\label{ENERGYestAuB}\\
&& =
{\textstyle{\frac{N}{N-2}}} 
\min_{\psi^\npow} \sum\sum_{\hskip-.7truecm {1\leq k < l\leq N-1}} 
\big\langle U^{(N)}_{k,l}\big\rangle_{\psi^{\npow}}.
\nonumber
\end{eqnarray}
	The first equality in \Ref{ENERGYestAuB} is just definition \Ref{qmEnullCOULOMB} and identity
\Ref{QformCOULrewriteDOUBLEsum}.
	For the first inequality in \Ref{ENERGYestAuB} we used identity \Ref{QformCOULgraphID} and the
fact that the minimum of a sum is never lesser than the sum of the minima; actually, this inequality 
is in general strict.
	For the second inequality in \Ref{ENERGYestAuB} we used that the to-be-minimized double sums before
that ``$\geq$'' symbol each involve only expectations computed with an $N-1$ point marginal of 
$h^{-3N} |\widehat{\psi}^{(\Npow)}_\hbar|^2 |\psi^{(\Npow)}|^2$, which can be written as
averages of conditional expectations --- conditioning is on the  $(\pV_n,\qV_n)$ variables of $\psi^{(\Npow)}$ ---
and the inequality results when the conditioning is relaxed; 
incidentally, since we do not impose any symmetry on the various $\psi$, the inequality 
symbol ``$\geq$'' can actually be replaced by the equality sign ``$=$'' (just tensor multiply each $N-1$ point
minimizing wave function with any nice 1 point wave function in the respective $n$-th variables), but if 
bosonic (or fermionic) symmetry is imposed, then the ``$\geq$'' is generally even a ``$>$.''
	For the final equality we used the permutation symmetry of $U^{(N)}_{k,l}$.

	Recalling \Ref{HamCOULrewrite} and \Ref{HamCOULpairTERM}, and letting $\widetilde\psi^{\npow}_{\Bose}$ 
denote the normalized minimizer (which exists!) of $\sum\sum\big\langle U^{(N)}_{k,l}\big\rangle_{\psi^{\npow}}$, 
with the double sum running over ${{1\leq k < l\leq N-1}}$, 
the last expression in \Ref{ENERGYestAuB} can be recast as follows,
\begin{eqnarray} 
\min_{\psi^\npow} \sum\sum_{\hskip-.7truecm {1\leq k < l\leq N-1}} 
\big\langle U^{(N)}_{k,l}\big\rangle_{\psi^{\npow}}
\hskip-0.6truecm&& =
\left\langle H^{(N-1)}_{\Coul}\right\rangle_{\widetilde\psi^{\npow}_{\Bose}} \!-
\label{ENERGYestAuBrewrite}\\
&&\qquad
{\textstyle{\frac{1}{(N-1)(N-2)}}} 
\sum\sum_{\hskip-.8truecm {1\leq k < l\leq N-1}}\!
\left\langle H^{(1)}_{\Coul,\,k} + H^{(1)}_{\Coul,\,l}\right\rangle_{\widetilde\psi^{\npow}_{\Bose}},
\nonumber
\end{eqnarray}
where 
\begin{equation}
H^{(1)}_{\Coul} := 
{\textstyle{\frac{1}{2m}}}|\pV|^2 - z^2e^2 \textstyle{{{\frac{1}{|\qV|}}}}
\label{hydroHamCoul}
\end{equation}
is a familiar Hydrogen-type Hamiltonian, and $H^{(1)}_{\Coul,\,k}$ and $H^{(1)}_{\Coul,\,l}$ indicate
that \Ref{hydroHamCoul} is expressed in the $k$-th and $l$-th particle's variables, respectively.
	To handle the Hydrogen-like contributions in the last line of \Ref{ENERGYestAuB} we
use the virial theorem, which for $N>2$ furnishes the identity
\begin{eqnarray}
	\hskip-1truecm
-{\textstyle{\frac{1}{N-2}}} \hskip-0.5truecm&&
\sum\sum_{\hskip-.8truecm {1\leq k < l\leq N-1}}\!
\left\langle H^{(1)}_{\Coul,\,k} + H^{(1)}_{\Coul,\,l}\right\rangle_{\widetilde\psi^{\npow}_{\Bose}}
=\nonumber\\
&&\hskip+3truecm
\bigl\langle H^{(N-1)}_{\Coul}\bigr\rangle_{\widetilde\psi^{\npow}_{\Bose}} +
{\textstyle{\frac{1}{N-2}}} z^2e^2\, \funcI^{(N-1)}(\widetilde\psi^{\npow}_{\Bose}) ,
\label{virialCOUL}
\end{eqnarray}
and since $\funcI^{(N-1)}(\widetilde\psi^{\npow}_{\Bose}) > 0$, we obtain the estimate
\begin{equation}
-{\textstyle{\frac{1}{N-2}}} 
\sum\sum_{\hskip-.8truecm {1\leq k < l\leq N-1}}\!
\left\langle H^{(1)}_{\Coul,\,k} + H^{(1)}_{\Coul,\,l}\right\rangle_{\widetilde\psi^{\npow}_{\Bose}}
>
\bigl\langle H^{(N-1)}_{\Coul}\bigr\rangle_{\widetilde\psi^{\npow}_{\Bose}} .
\label{virialCOULest}
\end{equation}	
	Estimates \Ref{ENERGYestAuB} together with identities \Ref{ENERGYestAuBrewrite} and \Ref{virialCOUL}
and the estimate \Ref{virialCOULest}, plus an obvious inequality, now give, for $N>2$,
\begin{eqnarray}
\cE_{\Coul}^{\Bose}(N)
\hskip-0.6truecm
&&>
{\textstyle{\frac{N^2}{(N-1)(N-2)}}} 
\bigl\langle H^{(N-1)}_{\Coul}\bigr\rangle_{\widetilde\psi^{\npow}_{\Bose}}
\nonumber\\
&&\geq
{\textstyle{\frac{N^2}{(N-1)(N-2)}}} 
\min_{\psi^{\npow}}\langle H^{(N-1)}_{\Coul}\rangle_{\psi^{\npow}}
\label{EofNvsEofNminusOneCoulomb}\\
&&
=
{\textstyle{\frac{N^2}{(N-1)(N-2)}}} 
\cE_{\Coul}^{\Bose}(N-1).
\nonumber
\end{eqnarray}
	Finally, dividing \Ref{EofNvsEofNminusOneCoulomb} by $N^2(N-1)$ yields, for $N>2$,
\begin{equation}
{\textstyle{\frac{1}{N^2(N-1)}}} \cE_{\Coul}^{\Bose}(N)
>
{\textstyle{\frac{1}{(N-1)^2(N-2)}}} \cE_{\Coul}^{\Bose}(N-1)
\end{equation}
and the proof of the monotonic increase of the map $N\mapsto\cE_{\Coul}^{\Bose}(N)/\cP_{\Coul}(N)$,
defined for $N\geq 2$, is complete.~\qed
\subsection{Proof of Proposition \ref{prop:monoUPb}}
\vskip-.3truecm
\noindent
	Up to where the virial identity is needed the proof of  
Proposition \ref{prop:monoUPb} follows verbatim the proof of Proposition \ref{prop:monoUPa}.
	The different order-three polynomials in Propositions \ref{prop:monoUPa} and
\ref{prop:monoUPb} are the result of necessarily different ``end games.''

	Thus, we first rewrite the quadratic form $\funcQ^{(N)}_{\Newt\!,M}(\psi^{(\Npow)})$ into the more convenient
format of a quasi-classical expectation functional, 
\begin{equation}
\funcQ^{(N)}_{\Newt\!,M}(\psi^{(\Npow)}) 
=
\iint H^{(N)}_{\Newt\!,M}h^{-3N} |\widehat{\psi}^{(\Npow)}_\hbar|^2 |\psi^{(\Npow)}|^2d^{^{3N}}\!\!\!p d^{^{3N}}\!\!\!q
=:\big\langle H^{(N)}_{\Newt\!,M}\big\rangle_{\psi^{(\Npow)}}
\label{QformNEWTrewrite}
\end{equation}	
where $H^{(N)}_{\Newt\!,M}(\pV_1,...,\qV_{N})$ is again the {\it classical} Hamiltonian with Newton interactions, 
formally also given by \Ref{HamiltonianSTAR} but now with $\pV_k\in\Rset^3$.
	Once again following Fisher and Ruelle \cite{FisherRuelle}, Dyson and Lenard \cite{DysonLenard}, and
L\`evy-Leblond \cite{LevyLeblond}, we rewrite the Newton Hamiltonian \Ref{HamiltonianSTAR} as a double sum,
\begin{equation}
H^{(N)}_{\Newt\!,M}
\equiv
\sum\sum_{\hskip-.7truecm  1 \leq k < l \leq N} U^{(N)}_{k,l}, 
\label{HamNEWTrewrite}
\end{equation}

\vskip-.3truecm
\noindent
but now
\begin{equation}
U^{(N)}_{k,l}
:=
{\textstyle{\frac{1}{2m(N-1)}}}\left( |\pV_k|^2 + |\pV_l|^2 \right) 
-  \textstyle{\frac{GMm}{N-1}} \left({{\frac{1}{|\qV_k|}}}+{{\frac{1}{|\qV_l|}}}\right)
-Gm^2{\textstyle{\frac{1}{|\qV_k-\qV_j|}}}.
\label{HamNEWTpairTERM}
\end{equation}
	The $^{(N)}$ at $U^{(N)}_{k,l}$ reminds us of the explicit $N$ dependence at r.h.s.\Ref{HamNEWTpairTERM}.
	With \Ref{HamNEWTrewrite} the quadratic form alias expectation functional 
\Ref{QformNEWTrewrite} becomes 
\begin{equation}	
\funcQ^{(N)}_{\Newt\!,M}(\psi^{(\Npow)})
\equiv
\sum\sum_{\hskip-.7truecm  1 \leq k < l \leq N} \big\langle U^{(N)}_{k,l}\big\rangle_{\psi^{(\Npow)}},
\label{QformNEWTrewriteDOUBLEsum}
\end{equation}	
and as before, for $N>2$ we have the identity
\begin{equation}
\sum\sum_{\hskip-.6truecm  1 \leq k < l \leq N} \big\langle U^{(N)}_{k,l}\big\rangle_{\psi^{(\Npow)}}
=
{\textstyle{\frac{1}{N-2}}}\!\!\sum_{1\leq n \leq N}
\sum\sum_{\hskip-.6truecm \genfrac{}{}{0pt}{}{1\leq k < l\leq N}{ k\neq n\neq l} } 
\big\langle U^{(N)}_{k,l}\big\rangle_{\psi^{(\Npow)}} .
\label{QformNEWTgraphID}
\end{equation}	
	Note that \Ref{QformNEWTgraphID} holds without assuming any particular symmetry of $\psi^{(\Npow)}$.

	We now start our estimates.
	Writing $\min_{\psi^{(\Npow)}}$ for the minimum over 
the subset of ${D}_\funcQ^{(N)}$ satisfying $\|\psi^{(\Npow)}\|_{\Lsp^2(\Rset^{3N})}=1$, for $N>2$ we find
\begin{equation}
\hskip-0.8truecm
\cE_{\Newt}^{\Bose}(N;M)
=
\min_{\psi^{(\Npow)}}
\sum\sum_{\hskip-.6truecm 1\leq k< l\leq N} \big\langle U^{(N)}_{k,l}\big\rangle_{\psi^{(\Npow)}}
\geq
{\textstyle{\frac{N}{N-2}}} 
\min_{\psi^\npow} \sum\sum_{\hskip-.7truecm {1\leq k < l\leq N-1}}\!\!\!
\big\langle U^{(N)}_{k,l}\big\rangle_{\psi^{\npow}}.
\label{ENERGYestAuBnew}
\end{equation}
	All the steps to get \Ref{ENERGYestAuBnew} are identical to the 
corresponding steps which yield formula \Ref{ENERGYestAuB}.

	The last expression can be recast with the help of elementary algebra, thus
\begin{eqnarray} 
\min_{\psi^\npow} \sum\sum_{\hskip-.7truecm {1\leq k < l\leq N-1}} 
\big\langle U^{(N)}_{k,l}\big\rangle_{\psi^{\npow}}
\hskip-0.6truecm&& =
\left\langle\! H^{(N-1)}_{\Newt\!,M}\!\right\rangle_{\!{\widetilde\psi^{\npow}_{\Bose\!,M}}}
\!\!-
\label{EofNminONErewrite}\\
&&\qquad 
{\textstyle{\frac{1}{(N-1)(N-2)}}} 
\sum\sum_{\hskip-.8truecm {1\leq k < l\leq N-1}}\!\!\!
\left\langle\!\! H^{(1,k)}_{\Newt\!,M} + H^{(1,l)}_{\Newt\!,M}\!\!\right\rangle_{\!{\widetilde\psi^{\npow}_{\Bose\!,M}}},
\nonumber
\end{eqnarray} 
where $\widetilde\psi^{\npow}_{\Bose\!,M}$ 
denotes the normalized minimizer of $\sum\sum\big\langle U^{(N)}_{k,l}\big\rangle_{\psi^{\npow}}$, 
the double sum running over ${{1\leq k < l\leq N-1}}$, and where
\begin{equation}
H^{(1)}_{\Newt\!,M} := 
{\textstyle{\frac{1}{2m}}}|\pV|^2 - GMm \textstyle{{{\frac{1}{|\qV|}}}}
\label{hydroHamNewt}
\end{equation}
is a familiar Hydrogen-type Hamiltonian, and the superscripts ${}^k$ and ${}^l$ at $H^{(1,\cdot)}_{\Newt\!,M}$ 
indicate that \Ref{hydroHamNewt} is expressed in the $k$-th and $l$-th particle's variables, respectively.
	Curiously, the identity \Ref{EofNminONErewrite} agrees exactly with the one in \Ref{ENERGYestAuBrewrite}; 
na\"{\i}vely one might have expected that the difference in the $N$-dependence of the central terms, viz. 
$Nz^2e^2$ vs. $GMm$, would already show itself at this point, but it does not. 
	Be that as it may, the strict similarity between the proofs of Propositions \ref{prop:monoUPa}
and \ref{prop:monoUPb} ends here.

	Namely, while we will also use the virial theorem to handle the Hydrogen-like terms
in the last line of \Ref{ENERGYestAuBnew}, we now first recast these terms as follows,
\begin{eqnarray}
\hskip-1truecm
{\textstyle{\frac{1}{N-2}}} \hskip-0.5truecm&&
\sum\sum_{\hskip-.8truecm {1\leq k < l\leq N-1}}\!
\left\langle H^{(1,k)}_{\Newt\!,M} + H^{(1,l)}_{\Newt\!,M}\right\rangle_{\widetilde\psi^{\npow}_{\Bose\!,M}}
=\nonumber\\
&&\hskip+3truecm
{\textstyle{\frac{\hbar^2}{2m}}} \funcK^{(N-1)}(\widetilde\psi^{\npow}_{\Bose\!,M})-GMm\, \funcC^{(N-1)}(\widetilde\psi^{\npow}_{\Bose\!,M}).
\label{PREPvirialNEWT}
\end{eqnarray}
	For $N>3$ the virial theorem now yields the identity
\begin{equation}
\hskip-1truecm
{\textstyle{\frac{\hbar^2}{2m}}} \funcK^{(N-1)}(\widetilde\psi^{\npow}_{\Bose\!,M})
=
-{\textstyle{\frac{N-1}{N-3}}} 
\bigl\langle H^{(N-1)}_{\Newt\!,M}\bigr\rangle_{\widetilde\psi^{\npow}_{\Bose\!,M}} 
+ {\textstyle{\frac{1}{N-3}}}GMm\, \funcC^{(N-1)}(\widetilde\psi^{\npow}_{\Bose\!,M}).
\label{virialNEWT}
\end{equation}
	Estimates \Ref{ENERGYestAuBnew} together with the identities \Ref{PREPvirialNEWT} and \Ref{virialNEWT},
plus the inequality $\funcC^{(N-1)}(\widetilde\psi^{\npow}_{\Bose\!,M})>0$, now give, for $N>3$,
\begin{eqnarray}
\cE_{\Newt}^{\Bose}(N;M)
\hskip-0.6truecm
&&\geq
{\textstyle{\frac{N}{(N-3)}}} 
\bigl\langle H^{(N-1)}_{\Newt}\bigr\rangle_{\widetilde\psi^{\npow}_{\Bose\!,M}}
\nonumber\\
&&\geq
{\textstyle{\frac{N}{(N-3)}}} 
\min_{\psi^{\npow}}\langle H^{(N-1)}_{\Newt}\rangle_{\psi^{\npow}}
\label{EofNvsEofNminusOneNewton}\\
&&
=
{\textstyle{\frac{N}{(N-3)}}} 
\cE_{\Newt}^{\Bose}(N-1;M);
\nonumber
\end{eqnarray}
the first inequality is strict if $N>4$.
	Dividing \Ref{EofNvsEofNminusOneNewton} by $N(N-1)(N-2)$ yields
\begin{equation}
{\textstyle{\frac{1}{N(N-1)(N-2)}}} \cE_{\Newt}^{\Bose}(N;M)
\geq
{\textstyle{\frac{1}{(N-1)(N-2)(N-3)}}} \cE_{\Newt}^{\Bose}(N-1;M)
\end{equation}
for $N\!>\!3$, with strict inequality for $N\!>\!4$.
	The proof of the monotonic increase of the map $N\mapsto\cE_{\Newt}^{\Bose}(N;M)/\cP_{\!\Newt}(N)$,
defined for $N\geq 3$, is complete.~\qed

\subsection{Proof of Proposition \ref{prop:monoUPc}}
\vskip-.3truecm
\noindent
	By the decomposition \Ref{HamiltonianSTARnullSPLIT}, the infimum of 
$\langle H^{(N)}_{\Newt\!,0}\rangle_{\psi^{\Npow}}$ equals the infimum of 
$\langle H^{(N-1)}_{\Newt\!,\rm int}\rangle_{\oli\psi{}^{\npow}}$, which with the help of 
\Ref{HamiltonianSTARintrinsic} can be shown to be a minimum; here, the overbar on ${\oli\psi{}^{\npow}}$
indicates dependence on $\{\oli\qV_k\}_{k=1}^{N-1}$ only.

	Writing $W^{(N)}_{k,l}$ for the summands on r.h.s.\Ref{HamiltonianSTARintrinsic}, we can apply
our strategy of proof of Propositions \ref{prop:monoUPa} and \ref{prop:monoUPb}.
	We just need to substitute $W^{(N)}_{k,l}$ for $U^{(N)}_{k,l}$ and
${\oli\psi{}^{\npow}}$ for ${\psi^{\Npow}}$, respectively ${\oli\psi{}^{\nnpow}}$ for ${\psi^{\npow}}$, 
 in \Ref{QformNEWTgraphID} and \Ref{ENERGYestAuBnew} (with $M=0$), rewrite as in
\Ref{EofNminONErewrite}, with $\langle W^{(N)}_{k,l}\rangle_{\widetilde{\oli\psi}{}^{\nnpow}_{\Bose}}$ 
in place of $\langle U^{(N)}_{k,l}\rangle_{\widetilde\psi^{\npow}_{\Bose\!,M}}$, and find
\begin{equation}
\hskip-0.6truecm
\cE_{\Newt}^{\Bose}(N;0)
\geq
{\textstyle{\frac{N}{N-2}}} 
\Big[\!\!\left\langle\! H^{(N-2)}_{\Newt\!,\rm int}\!\right\rangle_{\!{\widetilde{\oli\psi}{}^{\nnpow}_{\Bose}}}
\!\!-
{\textstyle{\frac{1}{N(N-1)}}} 
\sum\sum_{\hskip-.8truecm {1\leq k < l\leq N-1}}\!\!\!
	{\textstyle{\frac{1}{2m}}} 
\left\langle |\pV_j-\pV_k|^2 \right\rangle_{\!{\widetilde{\oli{\psi}}{}^{\nnpow}_{\Bose}}}
\!\Big]\!,
\label{ENERGYestAuBtrans}
\end{equation}
where $\widetilde{\oli\psi}{}^{\nnpow}_{\Bose}$ now denotes the normalized minimizer of 
$\sum\sum\big\langle W^{(N)}_{k,l}\big\rangle_{\oli\psi{}^{\nnpow}}$, 
the double sum running over ${{1\leq k < l\leq N-1}}$.
	At this point the virial theorem enters once again, but in contrast to the proofs of
Propositions \ref{prop:monoUPa} and \ref{prop:monoUPb} it here allows us to express 
r.h.s.\Ref{ENERGYestAuBtrans} entirely in terms of the expectation value of the intrinsic Hamiltonian, 
without producing any additional terms which would have to be estimated.
	So at the end of the day, our proof yields, for $N>2$,
\begin{equation}
\cE_{\Newt}^{\Bose}(N;0)
\geq
{\textstyle{\frac{N^2}{(N-1)(N-2)}}} 
\cE_{\Newt}^{\Bose}(N-1;0),
\label{EofNvsEofNminusONEnull}
\end{equation}
and dividing \Ref{EofNvsEofNminusONEnull} by $N^2(N-1)$ yields, for $N>2$,
\begin{equation}
{\textstyle{\frac{1}{N^2(N-1)}}} \cE_{\Newt}^{\Bose}(N;0)
\geq
{\textstyle{\frac{1}{(N-1)^2(N-2)}}} \cE_{\Newt}^{\Bose}(N-1;0).
\end{equation}
	The proof of the monotonic increase of the map $N\mapsto\cE_{\Newt}^{\Bose}(N;0)/\cP_{\Coul}(N)$,
defined for $N\geq 2$, is complete.~\qed

\vskip-.3truecm
\section{Further spin-offs: Hall--Post inequalities}
\vskip-.3truecm
\noindent
	Our Propositions \ref{prop:monoUPa}, \ref{prop:monoUPb}, and \ref{prop:monoUPc} are 
each equivalent to a statement that the $N$-body ground state energy $\cE(N)$ is bounded below by a specific 
$N$-dependent multiple of the $N-1$-body ground state energy $\cE(N-1)$; more precisely, when $N>N_0$ (with 
$N_0=2$ or $3$) then $\cE(N)\geq R(N)\cE(N-1)$, where $R(N)$ is some rational function of $N$.
	Here the so-compared $N$-body and $N-1$-body Hamiltonians \Ref{HamiltonianSTAR}, respectively 
\Ref{HamiltonianSTARintrinsic}, of our bosonic gravitational systems differ \emph{only} in their number 
of particles, and the Hamiltonians \Ref{HamiltonianATOM} of the bosonic atoms
differ \emph{only} in the number of their ``bosonic electrons'' and the corresponding charge of the atomic nucleus. 
	Within the adapted approximations (neglecting: relativity, spin degrees of freedom, nuclear motion, etc.)
our Propositions are therefore statements about sequences of systems as nature would supply them.
	These monotonicity results seem not to have been known before.

	Inspection of our proofs of  Propositions \ref{prop:monoUPa}, \ref{prop:monoUPb}, and 
\ref{prop:monoUPc} reveals that these proofs establish also technically 
somewhat stronger lower bounds for $\cE(N)$ in terms of a specific $N$-dependent multiple of the 
$N-1$-body ground state energy $\cE^\prime(N-1)$ of an $N-1$-body system with 
\emph{suitably rescaled coupling constants}.
	Such inequalities are known as Hall--Post inequalities; cf. \cite{KhareRichard}.
	Indeed, our inequalities \Ref{ENERGYestAuB}, \Ref{ENERGYestAuBnew} and \Ref{ENERGYestAuBtrans} 
in this paper are identical, in essence if not in appearance, to Hall--Post inequalities for our Hamiltonians. 
	Since the masses and charges of the various ``elementary'' particles of nature cannot 
be rescaled, nor can the ``constants of nature,'' these intermediate inequalities (``intermediate'' regarding
proving our propositions) are therefore \emph{generally not} statements about sequences of systems which nature could 
supply.

	There is (at least) one possible exception, though, to what we just wrote.
	Namely, in a certain quantum-mechanical approximation to QCD in which a 
baryon is made of an $N$-quark color-singlet state \cite{AderETal} the Hamiltonian has a factor $1/(N-1)$ 
in front of the pair interaction potential, and in this case the Hall--Post inequality relating the $N$- and
the $N-1$-body systems is precisely an inequality between the ground state energies of the $N$-quark and the 
$N-1$-quark Hamiltonians. 
	So in this case the Hall--Post inequality itself produces a monotonicity result for ``physical'' (i.e.,
according to that model) baryon masses $M(N)$, viz. the sequence $N\mapsto M(N)/N$ is monotonic increasing --- 
see (2.7) in  \cite{AderETal}. 
	The proof of (2.7) in \cite{AderETal} would not satisfy a mathematician, but it is ``morally correct'' and 
can easily be made rigorous (for a large class of pair-interaction potentials).
	Our strategy of proof is rigorous and produces the monotonicity result of \cite{AderETal}
in ``step one'' (NB: the virial theorem is not needed with $1/(N-1)$-rescaled pair interactions).
	Incidentally, the result of \cite{AderETal} is meant for fermionic quarks, but it holds for Bosons as well 
(see also our concluding remarks in the next section).

	The monotonicity result of \cite{AderETal} has spin-offs  analogous to our corollaries, not noted in 
\cite{AderETal}.
	Thus, for ``bosonic quarks'' an upper bound on the quark-specific baryon mass $M(N)/N$ as defined
by the approximation to QCD of \cite{AderETal} follows easily from the Hartree approximation, which 
together with the monotonic increase implies that the sequence $N\mapsto M(N)/N$ converges to a nontrivial limit.
	For fermionic quarks $N\mapsto M(N)/N$ is unbounded.

	Hall-Post inequalities between the ground state energies of $N$-body and $N-K$-body systems with
$K=1,2,...,N-2$ and appropriately rescaled coupling constants were first established in \cite{HallA}, picking 
up on earlier work in \cite{PostA}, \cite{HallPost} where $K=N-2$.
	However, the gist of the Hall--Post type proofs of the so-named inequalities is quite different from ours 
and relies heavily on the symmetry (or antisymmetry) of the wave function, which implies (in self-explanatory 
notation) that for each $N$ and each $\psi{}^{(\Npow)}$ all the 
$\big\langle |\pV|^2_{k}\big\rangle_{\psi{}^{(\Npow)}}$ have a common value, all the
$\big\langle V_{k,l}\big\rangle_{\psi{}^{(\Npow)}}$ have a common value, etc.
	By contrast, our proof of inequalities \Ref{ENERGYestAuB}, \Ref{ENERGYestAuBnew}, and 
\Ref{ENERGYestAuBtrans} in this paper does not make use of any symmetry of the wave functions and 
works equally well when the minimization is carried out over some subset of completely unsymmetric wave functions, 
should the demand arise.
	Moreover, with wave functions replaced by classical configurations, our technique handles also
the classical ground state problems with pair interactions $V(\qV_k,\qV_l)$ which are bounded below.
	Typically the $V(\qV_k,\qV_l)$ in a classical ground state configuration have \emph{no} 
common value, so that the Hall--Post strategy would fail to prove Proposition 1 in \cite{KieJSPa}. 

	To summarize, our proofs can be characterized in a nutshell by saying that their basic ingredients 
are the Hall--Post inequality and the virial identity for the respective Hamiltonian under study, 
plus some obvious positivity inequality --- except that we did not start from any Hall--Post inequality
but instead obtained the relevant inequality from scratch with a more flexible type of proof which does not
utilize any symmetry of the wave functions.
\section{Concluding remarks}
\noindent
	Our Propositions \ref{prop:monoUPa}, \ref{prop:monoUPb}, and \ref{prop:monoUPc} are
statements about the bosonic ground state energies of atoms and stars. 
	However, since we have not used any particular symmetry of $\psi^{(\Npow)}$ in our estimates,
Propositions \ref{prop:monoUPa}, \ref{prop:monoUPb}, and \ref{prop:monoUPc} and their proofs hold verbatim 
also for fermionic ground state energies $\cE_{\Coul}^{\Ferm}(N)$ and $\cE_{\Newt}^{\Ferm}(N;M)$, 
with $M\geq 0$, obtained by minimizing only over the anti-symmetric subspace of  $D^{(N)}_\funcQ$.
	So we conclude that also $N\mapsto\cE_{\Coul}^{\Ferm}(N)/\cP_{\Coul}(N)$ 
and $N\mapsto\cE_{\Newt}^{\Ferm}(N;M)/\cP_{\!\Newt}(N)$ and also
$N\mapsto\cE_{\Newt}^{\Ferm}(N;0)/\cP_{\Coul}(N)$ are monotonic increasing.
	Alas, Propositions \ref{prop:monoUPa},  \ref{prop:monoUPb}, and \ref{prop:monoUPc} are considerably 
less interesting for the ground state energies of fermionic atoms and stars than for their bosonic counterparts.

	Indeed, since $\cE_{\Coul}^{\Ferm}(N) \asymp -CN^{7/3}$ for large $N$ \cite{Lieb}, \cite{Thirring},
the monotonicity of $N\mapsto\cE_{\Coul}^{\Ferm}(N)/\cP_{\Coul}(N)$ is far from optimal.
	An optimal polynomial monotonicity result for the ground state energies of fermionic atoms would state
that $N\mapsto\cE_{\Coul}^{\Ferm}(N)/\cP_{\Coul}^{\Ferm}(N^{1/3})$ is monotonic increasing, 
where $\cP_{\Coul}^{\Ferm}(\,\cdot\,)$ is a polynomial of degree 7. 
	In comparison, by the upper bound $\cE_{\Coul}^{\Bose}(N)\!\leq\! -CN^3$ from Hartree theory one 
cannot improve our monotonicity result for the  bosonic atomic ground state energies to any lower leading 
power in $N$. 

	The same remarks apply \emph{mutatis mutandis} also to stars.
	Namely, since $\cE_{\Newt}^{\Ferm}(N;0) \asymp -CN^{7/3}$ for large $N$ (see \cite{Landau} for a formal 
argument and \cite{Thirring} for a proof), and presumably also $\cE_{\Newt}^{\Ferm}(N;M) \asymp -CN^{7/3}$ 
for large $N$, the monotonicity of $N\mapsto\cE_{\Newt}^{\Ferm}(N;M)/\cP_{\!\Newt}(N)$ and of
$N\mapsto\cE_{\Newt}^{\Ferm}(N;0)/\cP_{\Coul}(N)$ is far from optimal polynomial monotonicity 
for fermionic stars, namely that
$N\mapsto\cE_{\Newt}^{\Ferm}(N;M)/\cP_{\!\Newt\!,M}^{\Ferm}(N^{1/3})$ is monotonic increasing, where 
$\cP_{\!\Newt\!,M}^{\Ferm}(\,\cdot\,)$ is a polynomial of degree 7, indexed by $M\geq 0$.
	On the other hand, one cannot improve our monotonicity result for the bosonic stellar ground state energies 
to any smaller leading power in $N$, for $\cE_{\Newt}^{\Bose}(N;M)\leq -CN^3$ by Hartree theory.

	To prove the optimal polynomial monotonicity for the fermionic ground state energies of atoms and stars,
if possible at all, will require detailed input about the structure of the antisymmetric subspace of the form
domain. 
	Yet it may be hoped that the techniques developed in this paper will serve as an important stepping stone 
towards such fermionic proofs. 
\smallskip

\noindent
{\textbf{Acknowledgment}.} 
	This paper was written with support from the NSF under grant DMS-0807705.
	Any opinions expressed in this paper are entirely those of the author and not those of the NSF. 
	I thank Elliott H. Lieb for drawing my attention to the work of Post and collaborators, and
	Detlev Buchholz, Jerry Percus, Walter Thirring, and two anonymous referees for their helpful comments. 
\newpage

\end{document}